\documentclass[a4paper,fleqn,usenatbib,useAMS]{mnras}

\usepackage{graphicx}	
\usepackage{amsmath}	
\usepackage{amssymb}	
\usepackage{multicol}        
\usepackage{bm}		
\usepackage{pdflscape}	
\usepackage{subfigure}  
\usepackage{tabularx}	

\usepackage[T1]{fontenc}
\usepackage{ae,aecompl}

\usepackage{newtxtext,newtxmath}

\title[About 200-Year Cycle of Solar Activity]{About 200-Year Cycle of Solar Activity in the Mediaeval Korean Records and Reconstructions from Cosmogenic Radionuclides}

\author[K. Chol-jun]{Kim Chol-jun$^{1}$\thanks{Contact e-mail: \href{mailto:cj.kim@ryongnamsan.edu.kp}{cj.kim@ryongnamsan.edu.kp}}\thanks{Present address: Daesong District, Pyongyang, DPR Korea}
Kim Jik-su$^{2}$
\\
$^{1}$Physics Faculty, \text{Kim Il Sung} University, Daesong, Pyongyang, DPR Korea\\
$^{2}$Pyongyang Astronomical Observatory, Academy of Sciences of DPR Korea, Daesong, Pyongyang, DPR Korea}

\date{Last updated \today; in original form 8 August 2019}

\pubyear{2019}

\begin{document}
\label{firstpage}
\pagerange{\pageref{firstpage}--\pageref{lastpage}}
\maketitle

\begin{abstract}
We investigated the Korean records of naked-eye sunspot observations and found an implication of periodicity of about 200-year. Adding the Chinese records we showed that the historical naked-eye sunspot observations have the similar periodicity. Recent some works showed that there would be no intrinsic periodicities except 11-year cycle. We adopt the new approach called samplogram to test sampling stability of cycles in terms of power spectra and difference series and show that the Suess/de Vries cycle of about 207-year is a deterministic cycle of the stochastic solar activity. Also we show that occurrences of grand minimum are not necessarily expected with the Suess/de Vries cycle and it is possible to appear double or multiple grand maxima without a grand minimum within them.
\end{abstract}

\begin{keywords}
Sun: activity; methods: data analysis;
\end{keywords}




\section{Introduction}\label{sec:intro}
Korea has a longest history of astronomy in the world. Every historical chronicles in Korea accompanied records of astronomical phenomena.\citep{Samguk-Sagi, Koryo-Sa, Rizo-Sillok, Jungbo-munhon-bigo} We can gain some long-term variability of astronomical phenomena analyzing the Korean historical records. Solar activity is an instance that should be analyzed with data for long period. Though the records of naked-eye observations had irregularity and unknown quality, but the net tendency would have a good agreement with modern accurate measurements and there are some advantages that those data were measured directly(not estimated by some proxies) and covered long history.

Recently the reconstruction of the past solar activity is a very active field in solar science. Many studies\citep{Solanki2004, Steinhilber2012, Usoskin2016, Wu2018} of solar activity reconstructions have been done based on the cosmogenic radionuclides like $^{14}$C or $^{10}$Be. Individual dataset from both isotopes showed a lot similar but a bit different trends. The first consistent multi-proxy reconstruction of solar activity was proposed by \citet{Steinhilber2012}.\citet{Wu2018} also made a multi-proxy sunspot number(SSN) reconstruction in a composite way from a global $^{14}$C reconstruction(INTCAL09) and six $^{10}$Be datasets from Greenland(GRIP, NGRIP, Dye3) and Antarctica(EDML, DF, SP). Their reconstructions nearly cover the whole Holocene period of about 9 millennia. \citet{Beer2012} indicated that the only quantitative method of reconstructions of solar activity on centennial-millennial time scales is based on cosmogenic isotopes.

The long-term reconstruction of solar activity should follow by the analysis for the long-term solar cycles. \citet{Steinhilber2012} indicated the Suess/de Vries cycle($\sim$210-year), the Eddy cycle($\sim$1,000-year), and an unnamed cycle at approximately 350-year, as well as other less significant unnamed cycles at approximately 500 and 710-year from the power spectrum of their long-term reconstruction of solar activity. \citet{Abreu2012} showed a number of distinct periodicities, such as 88-year(Gleissberg), 104-year, 150-year, 208-year(Suess/de Vries), 506-year, 1000-year(Eddy), and 2200-year(Hallstatt) and discussed some of those cycles with relation to the planetary tides, though \citet{Poluianov2014} showed that their results were an artifact caused by the low sampling frequency. \citet{Usoskin2016} showed the so-called Hallstatt cycle of about 2400-year to be the most significant in long-term variability of solar activity. Those cycles, however, turned out to be spicular in power spectra and be intermittent in wavelet scalograms. Thus,  \citet{Cameron2013,Cameron2019} indicated that those cycles could not be distinguished from random spurious ones.

Cycles have appeared in old East Asian records of naked-eye observations of sunspots. \citet{Xu1990} analyzed his catalogue\citep{Wittmann1987} of naked-eye sunspot observations and concluded that beside 10.86-year period 212-year period is most significant cycle of solar activity. \citet{Ogurtsov2002} analyzed the data on ancient sunspot observations made by naked eye and said that the ancient oriental sunspot observations confirm the existence of the Suess/de Vries cycle($\sim$210-year) and give an evidence that century-scale solar variability has a wide frequency band(60-130 years). 

Analyzing the old Korean records of naked-eye sunspot observations we found a $\sim$200-year separation between concentrated periods of sunspot observations which is described in section~\ref{sec:korhis}. In comparison to modern reconstructions from cosmogenic isotopes, we cared of arguments of \citet{Cameron2019} that the cycles appear difficult to be proved as deterministic cycles in stochastic solar activity. We invented new methodology called samplogram that investigates how power of a peak changes in accordance to variation of sampling interval, which is shown in section~\ref{sec:stabcyc}, \ref{sec:disc}.

\section{Korean historical sunspot records and implications of about 200-year cycle from East Asian records}\label{sec:korhis}
Korea and East Asia including China have long or the longest official history even in astronomy. Investigating their historical records it is possible to find some long-term trends of astronomical phenomena like solar activity. The catalogues of sunspot from historical records had been compiled long ago. The most famous work is Kanda's catalogue which contains 142 items from Chinese, Korean and Japanese histories. The more complete catalogue was collected by \citet{Wittmann1987} which covers more than 18 centuries and includes more than 200 events. The Chinese historical records are being investigated even nowadays.\citep{Hayakawa2015, Hayakawa2017}

But the Korean sunspot records have been hardly investigated or referenced indirectly. Recently we have scanned old Korean historical chronicles\citep{Samguk-Sagi, Koryo-Sa, Rizo-Sillok, Jungbo-munhon-bigo} and compiled the data book for Korean astronomical observations.\citep{Jik-su2019} The dates of sunspot observations are listed in Table~\ref{tab:korssrec}. We could find that the sunspot records had been concentrated mainly in the late 12th century and late 14th century. Each intensive period had covered about 50 years, respectively. 

\begin{table*} 
\centering
 \caption{ The Korean records of sunspot observations.}
 \label{tab:korssrec}
 \begin{tabularx}{\textwidth}{lX}
  \hline
  Date&Original Text\\
  \hline
1151 Mar 21& On Kye-You$^a$ day of 3rd month in Ui-Jong$^b$ 5th year the sun had a black spot which was as big as a hen's egg$^c$.\\ 
1151 Mar 31 $\sim$ Apr 1&	On Kye-Mi day$^d$ the sun had a black spot which was as big as an egg. On Kap-Sin day the sun had a black spot which was as big as an egg.\\
1160 Sep 29&	On Kye-Yu day of 8th month in 14th year a black spot appeared in sun.\\
1171 Oct 20&	On Sin-Myo day of 8th month in Myong-Jong 1st year the sun had a black spot which was as big as a peach$^c$.\\
1171 Nov 16&	On Mu-In day of 10th month the sun had a black spot which was as big as a peach.\\
1183 Dec 4 $\sim$ 5&	On Kee-Myo day of 11th month in 13th year the sun had a black spot which lasted for 2 days.\\
1185 Feb 11&	On Kap-O day of 1st month in 15th year the sun had a black spot which was as big as a pear$^c$.\\
1185 Mar 27&	On Mu-O day of 2nd month the sun had a black spot which was as big as a pear.\\
1185 Apr 18 $\sim$ 19&	On Kyong-Ja day of 3rd month a black spot appeared in the sun. On Sin-Chuk day a black spot appeared in the sun.\\
1185 Nov 14&	On Kyong-O day of 10th month the sun had a black spot which was as big as a plum.\\
1198 Sep 30&	On Kye-Sa day of 8th month in Sin-Jong 1st year the sun had a black spot which was as big as a plum.$^e$\\
1200 Sep 19&	On Kye-Sa day of 8th month in 3rd year the sun had a black spot which was as big as a plum.\\
1201 Apr 6&	On Im-Za day of 3th month in 4th year the sun had a black spot which was as big as a plum.\\
1202 Aug 23&	On Byong-Za day of 8th month in 5th year the sun had a black spot which was as big as a pear.\\
1204 Feb 3 $\sim$ 5&	On the 1st Ul-Chuk day of 1st month in 7th year the sun had a black spot which was as big as a plum. It lasted for 3 days.\\
1258 Sep 15 $\sim$ 16&	On Kye-Sa day of 8th month in Ko-Jong 45th year the sun had a black spot which was as big as a hen's egg. The next day appeared a black spot again which looked like a human in shape.\\
1278 Aug 31&	On Kye-Hae day of 8th month in Chung-Ryol 4th year the sun had a black spot which was as big as a hen's egg.\\
1353 Apr 20&	On Kap-Sin day of 3th month in Kong-Min 2nd year the sun had no shine and a black spot appeared in it which keeps the same the next day.$^e$\\
1356 Mar 4 $\sim$ 5&	On Kap-Sin day of 3th month in 5th year the sun had no shine and a black spot appeared in it. On Ul-You day the sun had no shines and a black spot appeared in it.\\
1361 Mar 16 $\sim$ 19&	On Sin-Myo day of 2nd month in 10th year a black spot appeared in the sun which lasted for 4 days.\\
1362 Oct 5&	On Kee-Mee day of 9th month in 11th year a black spot appeared in the sun.\\
1371 Jan 2&	On Kyong-O day of 12th month in 19th year a black spot appeared in the sun.\\
1371 Sep 22&	On Kye-Sa day f of 9th month a black spot appeared in the sun.$^e$\\
1372 May 8&	On Im-O day of 4th month in 21th year a black spot appeared in the sun.\\
1373 Apr 26 $\sim$ 27&	On Ul-Hae day of 4th month in 22th year a black spot appeared in the sun which lasted for 2 days.\\
1373 Oct 23&	On Ul-Hae day of 10th month a black spot appeared in the sun.\\
1375 Mar 20 $\sim$ 21&	On Mu-Sin day of 2th month in Sin-U 1st year a black spot appeared in the sun. On Kee-You day a black spot appeared in the sun.\\
1381 Mar 23&	On Kye-Mi day of 2th month in 7th year a black spot appeared in the sun.\\
1382 Mar 9 $\sim$ 11&	On Kap-Sul day of 2th month in 8th year the sun had a black spot which was as big as a hen's egg and lasted for 3 days.\\
1387 Apr 15&	On Jong-Chuk day of  3th month in 13th year a black spot appeared in the sun.\\
1402 Nov 15&	On Kyong-O day of 10th month in Tae-Jong 2nd Im-O year a black spot appeared in the sun. \\
1556 Apr 17&	On Jong-Myo day of 3rd month  in Myong-Jong 11th Byong-Jin year in the midst of the sun there was a black spot as big as a hen's egg.\\
1604 Oct 24 $\sim$ 25&	On Kee-Myo day of 9th month in Seon-Jo 37th  Kap-Jin year leap in the center of the sun there was a black spot as big as a bird's egg in about to rising the sun. On Kyong-Jin day a black spot was in the sun which was as big as an egg in about to rising the sun.\\
1648 Jan 16&	On Mu-Za day of 12th month in In-Jo 25th Jong-Hae year in the center of the sun there was a black spot.\\
  \hline
  \end{tabularx}
\raggedright $^a$ Every day of month was called not by a number but by a word that means an animal.\\
$^b$ Ui-Jong means the king's legal name. Once the new one comes to the throne, the new legal name is attached to calling the calendar years which are renumbered since then.\\
$^c$ We cannot scale the size of expressions "pear," "egg," "plum," and "peach," which we only imagine as a spot of a few heliographic degrees. \\
$^d$ of the same month in the same year as above record.\\
$^e$ Data missed in \citet{Wittmann1987}.\\
$^f$ There was no Kye-Sa day in that month so it is dubious.\\
\end{table*}

Those periods belonged mainly to Koryo dynasty(AD 918-1392) which was the first unified dynasty in Korean history. In Koryo dynasty the astronomical observations were very systematic. We can find the sunspot records in later the Korean feudal dynasty(AD1392-1910), which, however, did not show such concentrations but look rather sporadic and sparse. Before Koryo dynasty we could not find any records yet. Both intensive periods mentioned above are separated by about 200 years and they lead us to think of about 200-year periodicity of solar activity. Post-Koryo records had been in late 16th or early 17th centuries which also are separated from the last intensive period(late 14th century) by 200 or 250 years. We see a few sunspot records between 1205 and 1352 which we call the sunspot gap. Then is this a physical gap in solar activity or an artificial gap due to broken observations or lost data? Are there any possibilities for the sunspot observations to be missed or lost due to other reasons such as wars, rebellions and so on? We scanned the political circumstances during Koryo dynasty. What must be said first is the fact that Koryo dynasty had maintained its rule over the territory in spite of that most of world and East Asia had been conquered by medieval Mongolians. It gave a favorable situation to astronomical observations. Actually during the sunspot gap there were the solar eclipse records almost every 2 or 3 years. So we could neglect the possibilities for the sunspot observations to be missed or lost by political reasons or due to a broken observational system.

\begin{figure}
\includegraphics[width=\columnwidth]{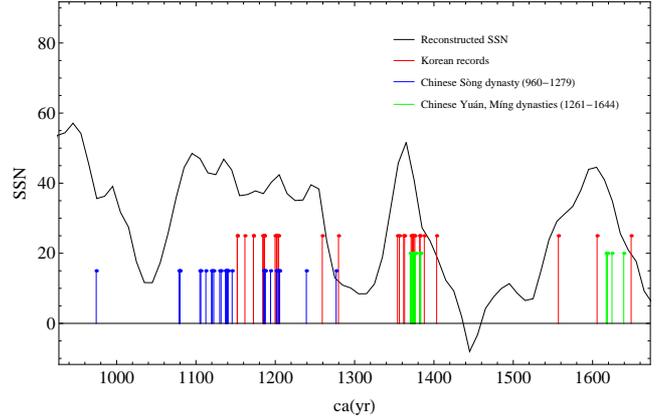}
 \caption{ \label{fig:hisssrec}The Korean and Chinese records of sunspot observation and reconstructed sunspot number data. We used the red long vertical bars for the Korean records, the blue short for Chinese S\`{o}ng dynasty(960-1279) and the green middle for Chinese Yu\'{a}n and M\'{i}ng dynasties(1261-1644). The solid line indicates reconstruction of sunspot number(SSN) from cosmogenic radionuclides. Height of vertical bars for naked-eye sunspot observations does not mean sunspot numbers.}

\end{figure}

We compare the Korean records with the contemporary Chinese records of the same times. The catalogue of \citet{Wittmann1987} has already included almost records in Table~\ref{tab:korssrec} that we have compiled. The Chinese records that \citet{Hayakawa2015, Hayakawa2017} had compiled included an intensive period of the 12th, late 14th and early 17th centuries. Figure~\ref{fig:hisssrec} shows records in the Korean and Hayakawa's pure Chinese catalogues. The Korean and pure Chinese records have little common dates which does not, however, give rise to any suspect of regularity of Korean observations. In fact, the catalogue of \citet{Wittmann1987} showed little common dates in several places and \citet{Hayakawa2015, Hayakawa2017} also showed the tables of solar activities which did not include any common date in several places. 

As said earlier, the records of naked-eye sunspot observation are irregular. Note that while we are more or less sure of sunspot occurrence when there is a record, the absence of the records does not imply the absence of sunspots. The sunspot record does reflect not only the solar activity, but also the frequency of observation and recording. But, mathematically speaking, the naked-eye sunspot observation records can be considered a kind of random sampling where the sampling interval varies randomly. The random sampling can show the intrinsic periodicity, though the precision is low, and it has some advantages like anti-aliasing\citep[see, for example,][]{Hajar2017}. Here we, however, would not like to conclude about a cycle from historical records.

The replenishment with Chinese historical data still shows about 200 or 250-year periodicity.(Figure~\ref{fig:hisssrec}) The same results had been obtained in previous works\citep[see][section~\ref{sec:intro}]{Xu1990, Ogurtsov2002}. Throughout the works about historical sunspot records we can assure that the naked-eye sunspot observations have a significant periodicity of $\sim$200-year. 

We are going to compare those East Asian historical records with recently reconstructed sunspot number(RSSN) dataset and justify our conjecture of about 200-year period. We compared the East Asian sunspot records with the RSSN time series\footnote{\url{https://www2.mps.mpg.de/projects/sun-climate/data/SN_composite.txt}} by \citet{Wu2018} in Figure~\ref{fig:hisssrec}. It is clear that the historical naked-eye sunspot observations have a good agreement with the RSSN. In 12th century the Korean and Chinese data have shown a wide intensive period. Recall that the apparent sunspot concentrated periods have shown about 200-year cycle.

\section{The Choice of Stochastic Stable Cycle in Long-term RSSN data}\label{sec:stabcyc}

In spite of implications of about 200-year cycle in historical and some reconstruction data, the recent works, however, suggested that the RSSN might have no intrinsic periodicities except well-known 11-year cycle. \citet{Cameron2019} wrote that the power spectrum of the RSSN from cosmogenic isotopes is consistent with a weakly nonlinear and noisy limit cycle with no intrinsic periodicities except that of the basic 11/22-year cycle. They showed that the so-called $\sim$90-year Gleissberg and $\sim$210-year Suess/de Vries cycles have narrow width in power spectrum of RSSN series so that those seem to be the byproducts of stochastic process of RSSN. Actually many candidates of cycles in RSSN appear spicular in power spectra and intermittent in wavelet scalograms so that it is difficult to differentiate them from stochastic spurious cycles. We try to find any periodicity in the RSSN series which is a stochastic stable, that is, a deterministic cycle.

As the previous authors, we have not gained the clear deterministic cycles in Fourier or wavelet analyses of the RSSN series. Recalling, however, about 200-year cycle of the naked-eye sunspot observations, we try to reconstruct the cycle from the reconstruction data.

First, we paid attention to the difference series from the RSSN series because if we consider the RSSN series a discrete stochastic process, their difference or increment series would represent the underlying dynamics. The RSSN data of \citet{Wu2018} is a decadal time series so the series represents sunspot numbers averaged over intervals of 10 years. The reconstructed total solar irradiance(RTSI) series\footnote{\url{https://www1.ncdc.noaa.gov/pub/data/paleo/climate_forcing/solar_variability/steinhilber2012.txt}} of \citet{Steinhilber2012} has interval of 22 year in time. We denote RSSN or RTSI of every data point by $X(i)$. Here $i$ means the index of an element in the series. Their difference series consist of difference between RSSN or RTSI values of adjacent points: 
\begin{align}
\Delta X(i)=X(i+1)-X(i)
\end{align}
Though the differences of sunspot numbers may not represent an independent physics, while they should have the same periodicity as the sunspot number series, they would have the mathematical significance to be analyzed. 

Second, we evaluated the correlation between the difference series of RSSN and the shifted(delayed or advanced) series itself by some time steps which is the autocorrelation of difference series as a function of time shift. The autocorrelation is evaluated as the normalized covariance:
\begin{align}\label{eq:ac}
R_{\Delta}(m)&=\frac{\langle\Delta X_1-\langle\Delta X_1\rangle \rangle\langle\Delta X_2-\langle\Delta X_2\rangle \rangle}{(V(\Delta X_1)V(\Delta X_2))^{1/2}},
\end{align}
\noindent where
\begin{align}
\Delta X_1&=\Delta X(1;;N-m), \nonumber\\
\Delta X_2&=\Delta X(m+1;;N), \nonumber
\end{align}
and $\Delta X(i)$ is the difference series of sunspot number, $i$ is the index of an element and $\Delta X(i;;j)$ is a set of elements numbered from $i$ to $j$. $\langle\rangle$ is the mean value of a set and $V()$ is the variance of it. $N$ is the total length of the series and $m$ stands for the shifted time steps.

\begin{figure}
\centering
\subfigure[]{
\includegraphics[width=\columnwidth]{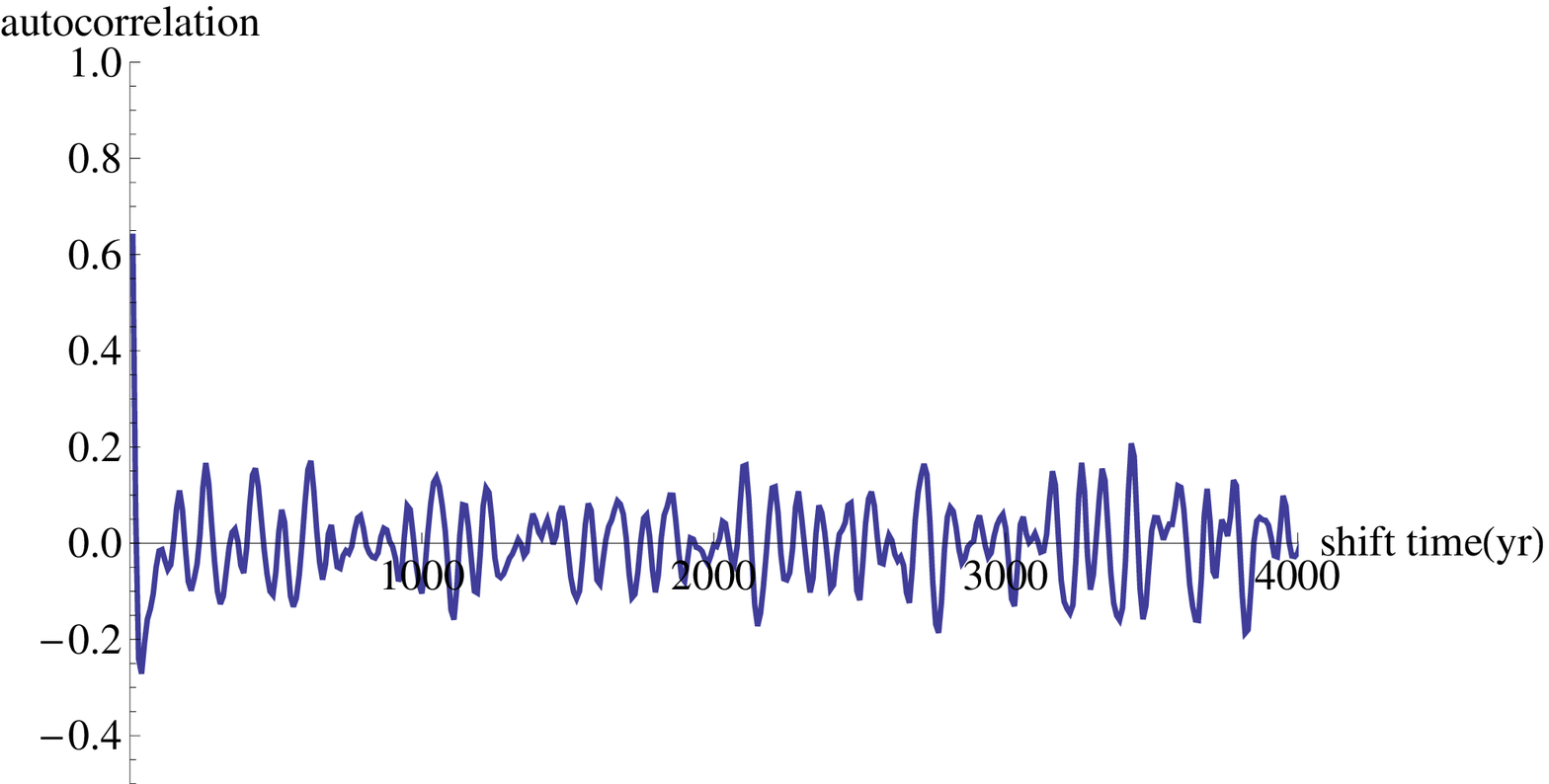}}
\subfigure[]{
\includegraphics[width=0.9\columnwidth]{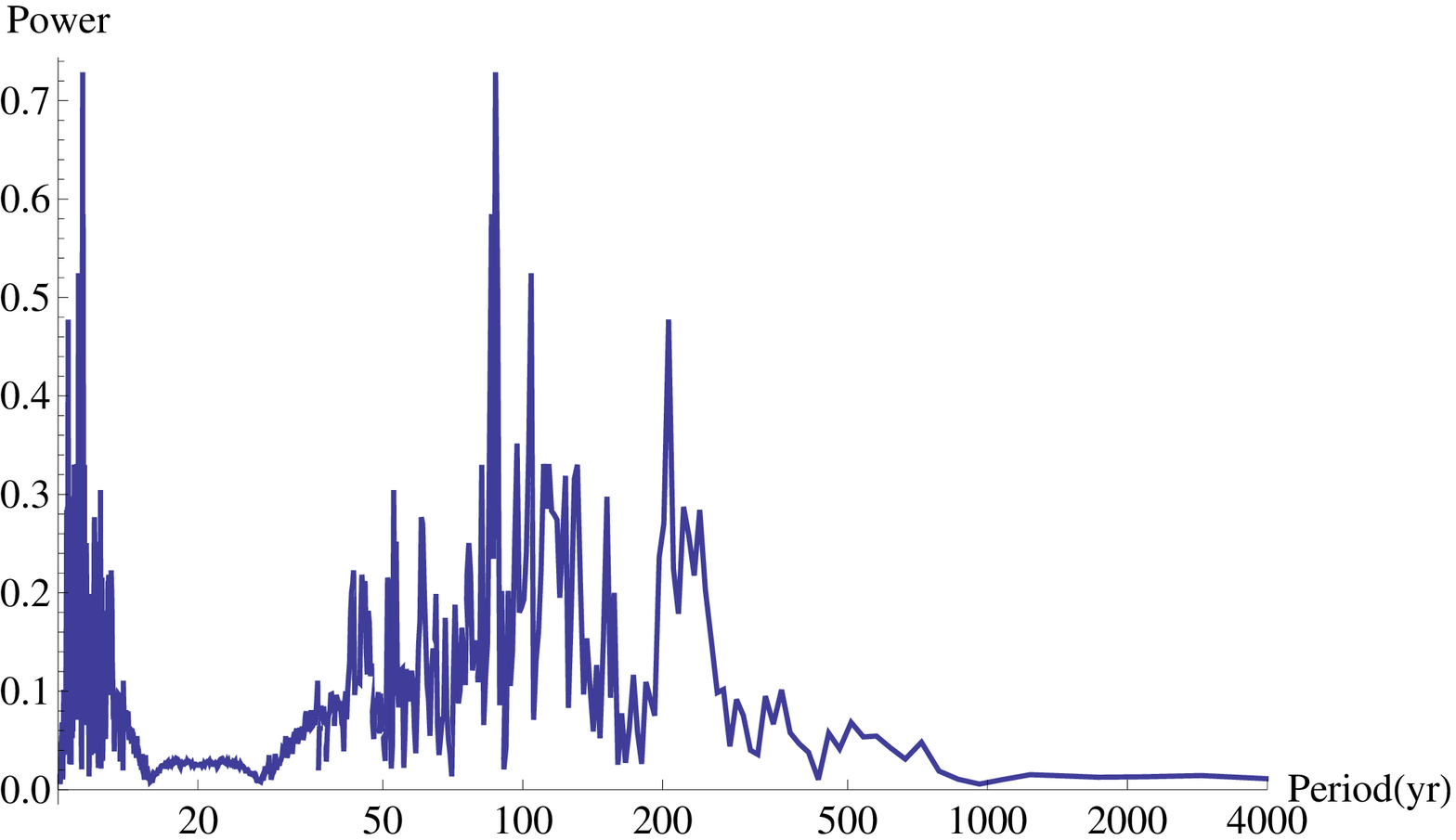}}
 \caption{\label{fig:acps} Autocorrelation of the difference series in terms of shift time(a) and its power spectrum(b).}
\end{figure}

\begin{figure}
\subfigure[]{
\includegraphics[width=\columnwidth]{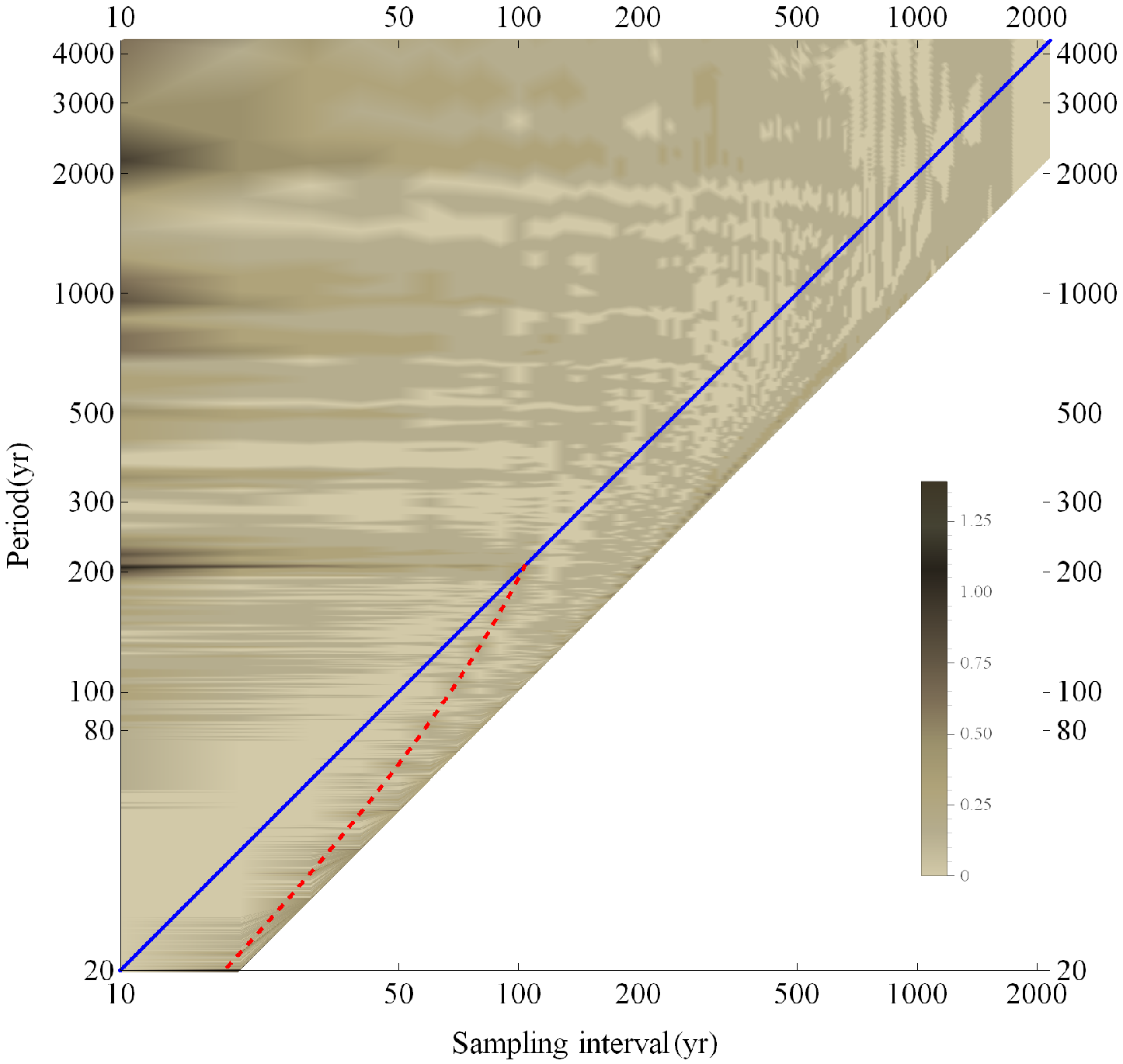}}
\subfigure[]{
\includegraphics[width=\columnwidth]{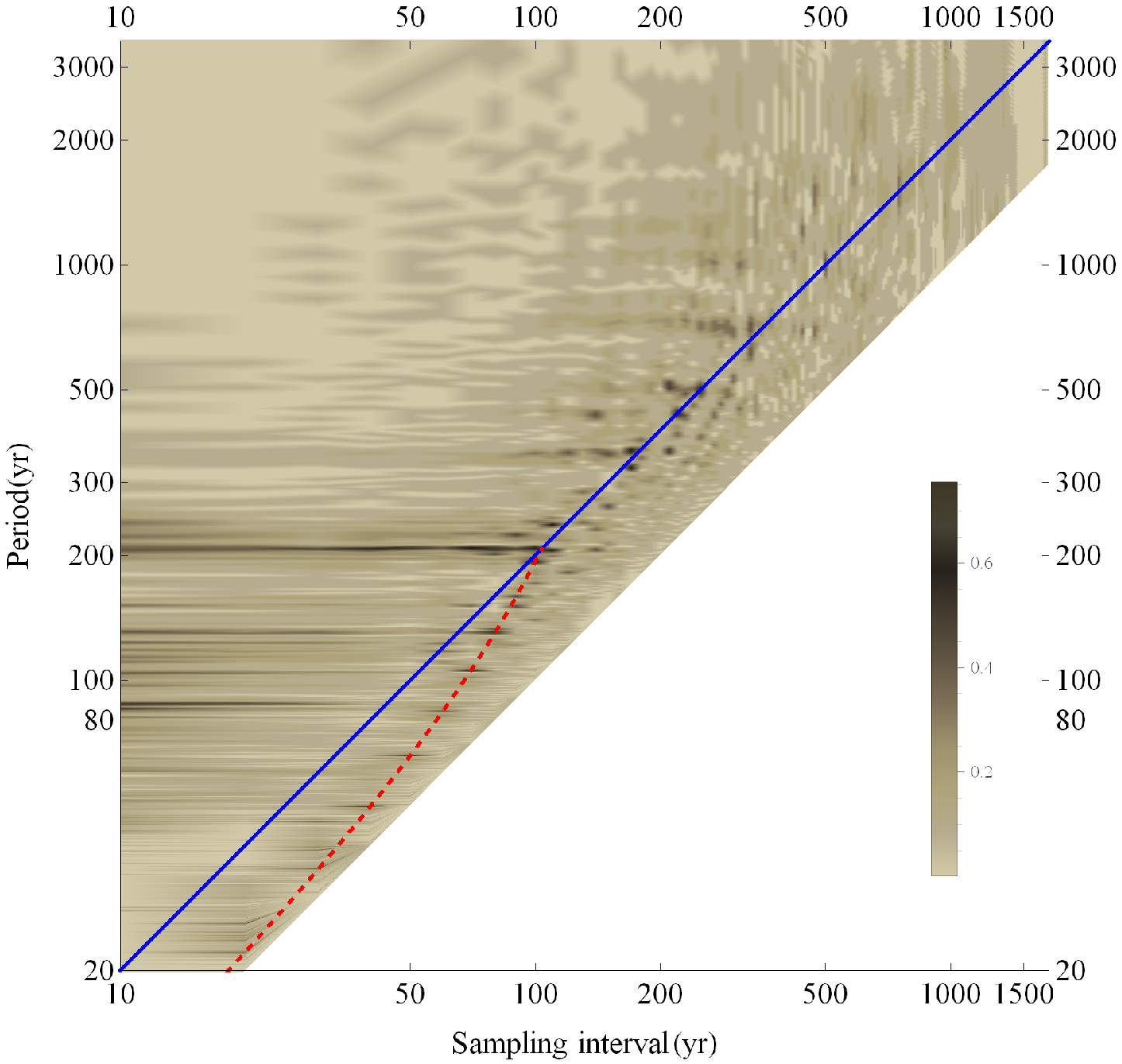}}
 \caption{\label{fig:samplogram} Behavior of power peaks of RSSN series(a) and its difference series(b) in varying sampling interval. The solid line implies the limit of the sampling theorem that means the sampling interval is half period of cycles. The lower boundary of the diagram implies the limit of Fourier analysis that means the sampling interval is equal to the period of cycles. The dashed line shows the track of aliasing peak of 207-year cycle. In (b) some "peaks" are seen to soar just before the Nyquist limit and decay beyond the limit.}
\end{figure}

Note that we used the RSSN of \citet{Wu2018} as mentioned above which is a decadal time series. Autocorrelation with shift time step of 1(that is, shift time is a decade) is 0.63. While increasing the shift time, autocorrelation drops rapidly to less than 0.2. But it never converges to zero, it is just oscillating. Period appears to be about 100 years. Figure~\ref{fig:acps} shows the autocorrelation of the difference series vs. shift time in years and the Fourier spectrum of the autocorrelation in terms of shift time, namely, the power spectrum of the difference series. The power spectrum has many peaks which are grouped mainly into three periods: 10-year, about 100-year, about 200-year. The first group of peaks looks to be related to the sampling interval which is a decade in our case. However, according to the sampling theorem, the region from 10 year to 20 year is the aliasing region so it is only a reflection of the region from 20 year to infinity. We keep, however, this region because this region means beyond sampling(Nyquist) limit later in Figure~\ref{fig:samplogram}, where behavior of peaks beyond the Nyquist limit is discussed. The second and third groups remind us the Gleissberg($\sim$90-year) and Suess/de Vries cycles($\sim$210-year). Note that there are no significant peaks of greater than 2000-year. Especially we could not find the so-called Hallstatt cycle of about 2400-year. The power spectrum is very plain in range corresponding to the very long-term cycles.

Third, we try to test the sampling stability as a stochastic stability of above cycles, that is, to determine whether those cycles may have the random peaks or the deterministic ones. In order to do that, we vary the sampling interval. While varying the sampling interval, the noisy peaks of power spectrum will vary randomly or decay rapidly. According to our simulation, the additive randomness to a simple harmonic signal causes the Fourier amplitude of the signal to vary randomly around the deterministic decay(see section~\ref{sec:disc}) while increasing sampling interval. On the other hand, \citet{Hajar2017} showed through their machine experiment that the noisy peaks decay more rapidly than the deterministic peaks in down sampling. We can say hardly about the general behavior of random peaks because we cannot expect all the types of random signal. We can expect surely only that the random peak should have no regular decay as the deterministic peaks or non-stability in down-sampling. 

We increase sampling interval and average the steps of RSSN series within each sampling interval: If we take a sampling step n, then sampling interval is $n\times\Delta t$, where $\Delta t$ is the time step of a reconstruction dataset so that, for example, $\Delta t$ is 10 year for \citet{Wu2018} and 22 year for \citet{Steinhilber2012} If $n$ is 3, evaluate mean value of subsets from elements numbered like $\lbrace1, 2, 3\rbrace, \lbrace4, 5, 6\rbrace$ and, in general, $\lbrace n\times(i - 1) + 1, \dots, n \times i \rbrace$ which will be the ith element of the new down-sampled(or coarser sampled) series. The difference series of the down-sampled series is obtained simply by getting the differences between adjacent elements. So the time step of the difference series is the same as that of the down-sampled series and the total number of elements is lowered by 1.

Then, we take the autocorrelation of the down-sampled(or coarser sampled) RSSN and apply the Fourier analysis to it as a function of shift time. We can see the results in Figure~\ref{fig:samplogram}(b). This figure shows that how the peak behaves in varying sampling interval. The lower boundary of the diagram implies the limit of power spectrum which gains cycles of period down to the sampling interval. We call this boundary the Fourier limit. Note that the solid line represents the limit of the sampling interval according to the sampling theorem which says that the sampling interval of a time series must be less than half of period of cyclic behavior of that series. We cannot expect any cycle of period shorter than the twice sampling interval. We call this the Nyquist limit which is half of the period of interest. Figure~\ref{fig:samplogram}(b) shows that only the significant peak of about 200-year is maintained just up to the Nyquist limit in varying the sampling interval. The miscellaneous periodicities of between 80- and 150-year decay in increasing sampling interval. As mentioned above, there are no significant long-term cycles of greater than 2000-year. The peaks of between 300- to 1000-year show the peculiar properties around the Nyquist limit that the peaks appear just before the limit and disappear after the limit.

We apply the same analysis to the RSSN series itself. Thus we plot power peaks of the original RSSN series itself vs. sampling interval.(see Figure~\ref{fig:samplogram}(a)) For the RSSN series, there are no significant peaks corresponding to the cycles of shorter than 200-year which shows a great contrast to the case of difference series of RSSN. Instead there appear several peaks of much long periods including the Hallstatt cycle. But those long periodicities decay with increasing the sampling interval. Every peak weakens in increasing the sampling interval. However, only the peak of about 200-year extends to the Nyquist limit. So we can conclude that the cycle of $\sim$200-year is a deterministic cycle in solar activity, which is just the Suess/de Vries cycle of 207-year according to our estimation. As we have mentioned above, if it were a random peak, its height would vary randomly with increasing sampling interval. However, as we can see, it keeps its peak just to the Nyquist limit, which says that the cycle is deterministic. 

We apply the above procedure to the reconstructed TSI series of \citet{Steinhilber2012} which has a time step of 22-year. The results are very similar, which justifies our inference. 

Last, we rearrange the grand maxima and grand minima in RSSN series in terms of the Suess/de Vries cycle. Since \citet{Eddy1976} had designated the Maunder minimum, several named grand minima and maxima had been allocated over the past millennium: the Oort minimum(11th century), the Wolf minimum(late 13th century), the Sp\"{o}rer minimum(mid-15th century), the Maunder minimum(late 17th century), the Dalton minimum(early 19th century) and the Mediaeval maximum(12th century), the Modern maximum(late 20th century). Those grand minima and maxima appeared also in the RSSN time series over past millennium so that RSSN has the cyclic rise and fall. Several studies argued that a new grand minimum will come at the beginning of this century\citep{Morner2015}. Such an expectation seems to have a basis on the idea that the grand minima may have periodicity of Suess/de Vries cycle. In the whole Holocene period, however, the periodicities couldn't be found in the raw RSSN series at a glance. 

\begin{figure}
\subfigure[]{
\includegraphics[width=\columnwidth]{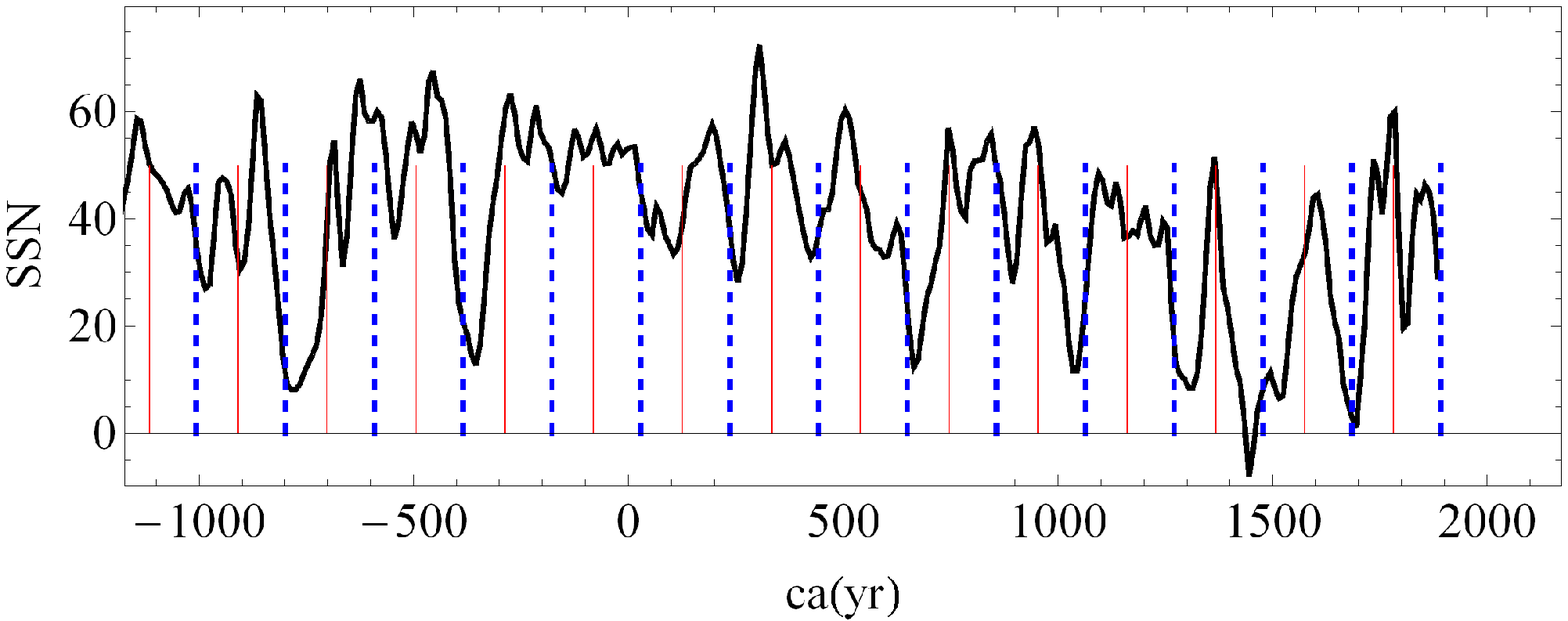}}
\subfigure[]{
\includegraphics[width=\columnwidth]{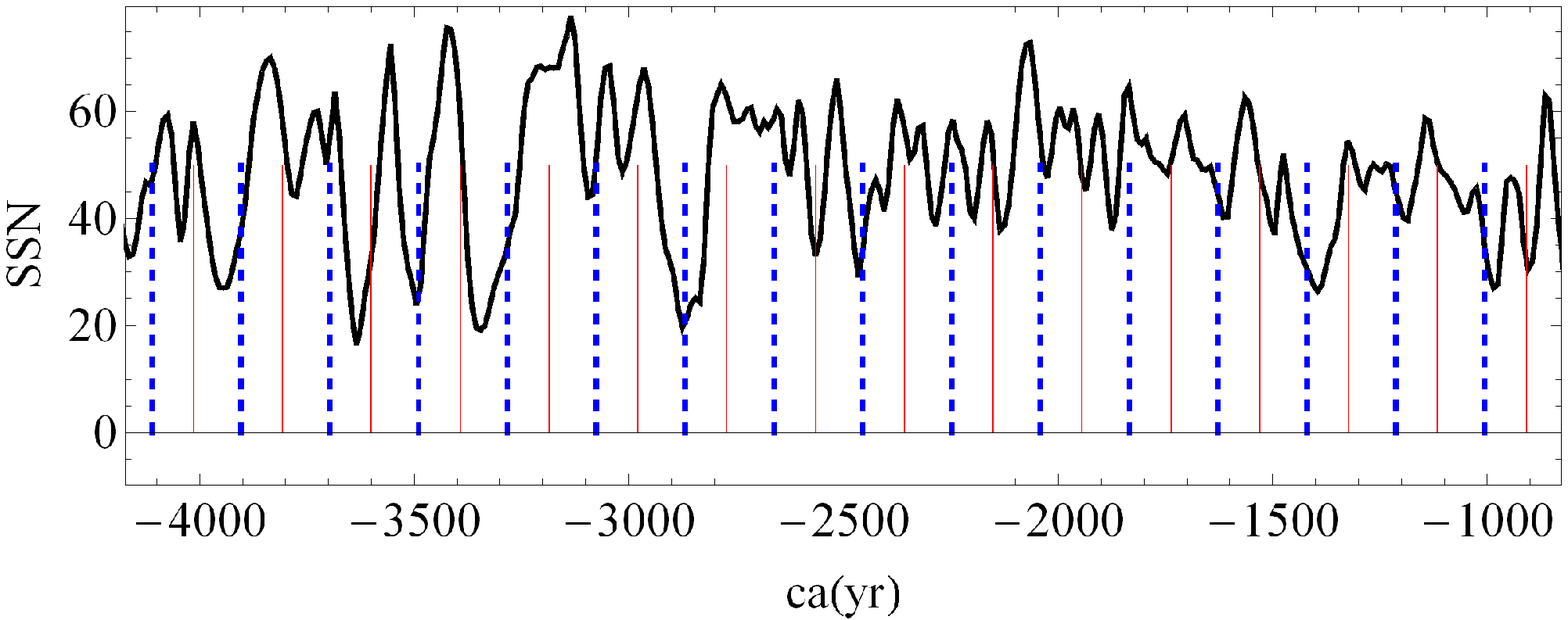}}
\subfigure[]{
\includegraphics[width=\columnwidth]{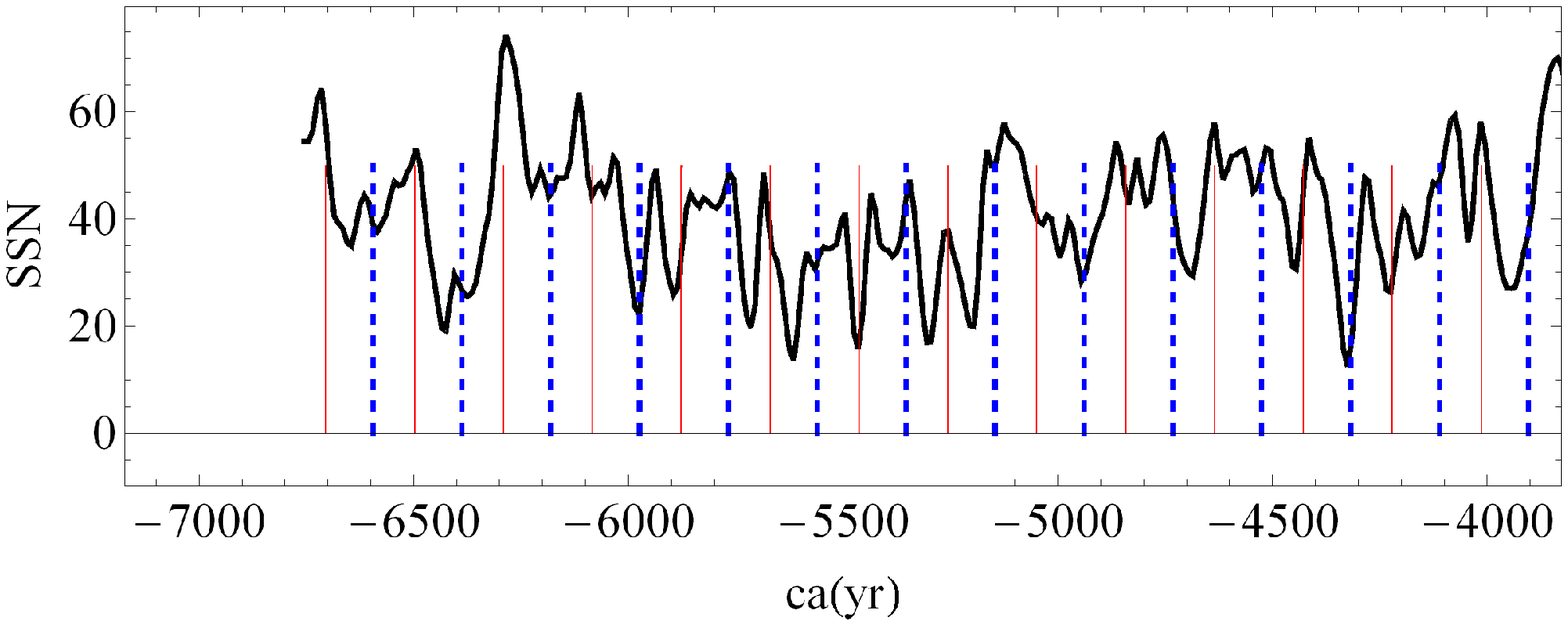}}
 \caption{\label{fig:RSSN200} The RSSN series with the Suess/de Vries cycle. Solid vertical bars span by the Suess/de Vries cycle maxima including 1989 grand maximum, while the dashed bars--minima including 1685 the Maunder minimum. It is remarkable that there are the double or multiple grand maxima without a deep grand minimum between adjacent grand maxima.}
\end{figure}

\citet{Inceoglu2015, Inceoglu2016} and \citet{Usoskin2016} defined the grand minima and grand maxima in RSSN series with some criteria such as duration and sunspot number. Among those grand minima and maxima it is possible to find the separations between both adjacent ones are close to 200 years, but those are intermittent. The periodicity of about 200-year can be more easily found in the lists of grand maxima rather than the lists of grand minima. We showed the RSSN series with ticks of 207-year periodicity including 1989 grand maximum and 1685 the Maunder minimum of solar activity in Figure~\ref{fig:RSSN200}. We could define the grand minimum and maximum only from the absolute values in RSSN series: grand maxima at sunspot number of around 60 and grand minima at around 20. If one neglect and hide vertical bars implying this Suess/de Vries cycle, it will be hard to find any periodicities. These bars could help you to imagine the Suess/de Vries cycle in the stochastic solar activity.

\begin{figure}
\centering
\subfigure[]{
\includegraphics[width=0.8\columnwidth]{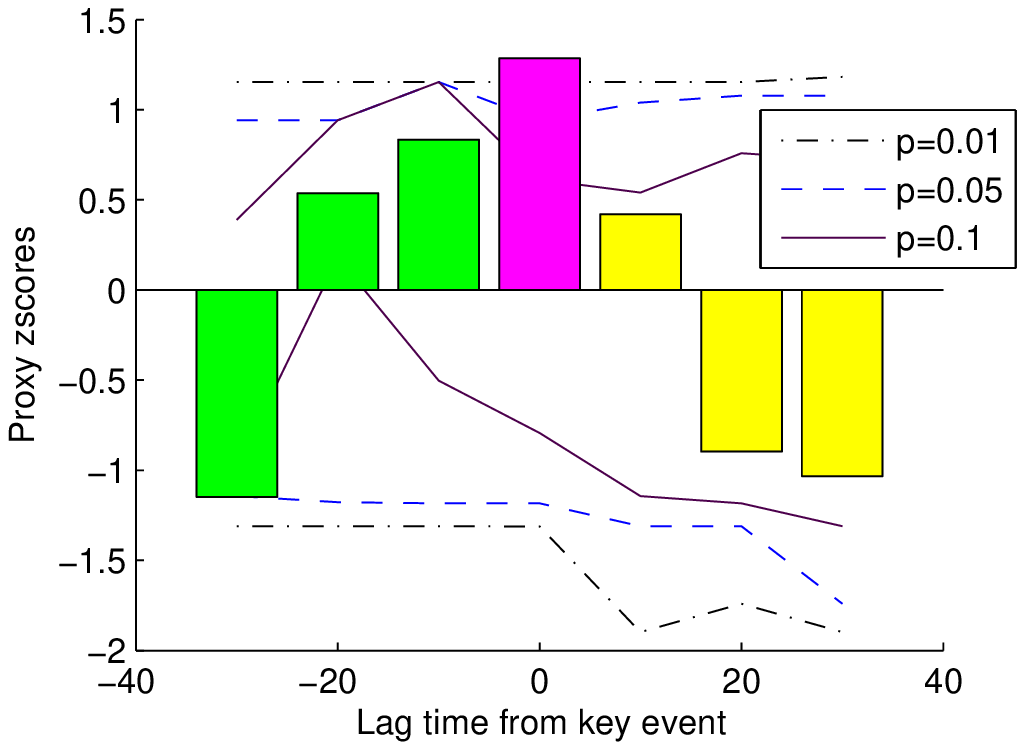}}
\subfigure[]{
\includegraphics[width=0.8\columnwidth]{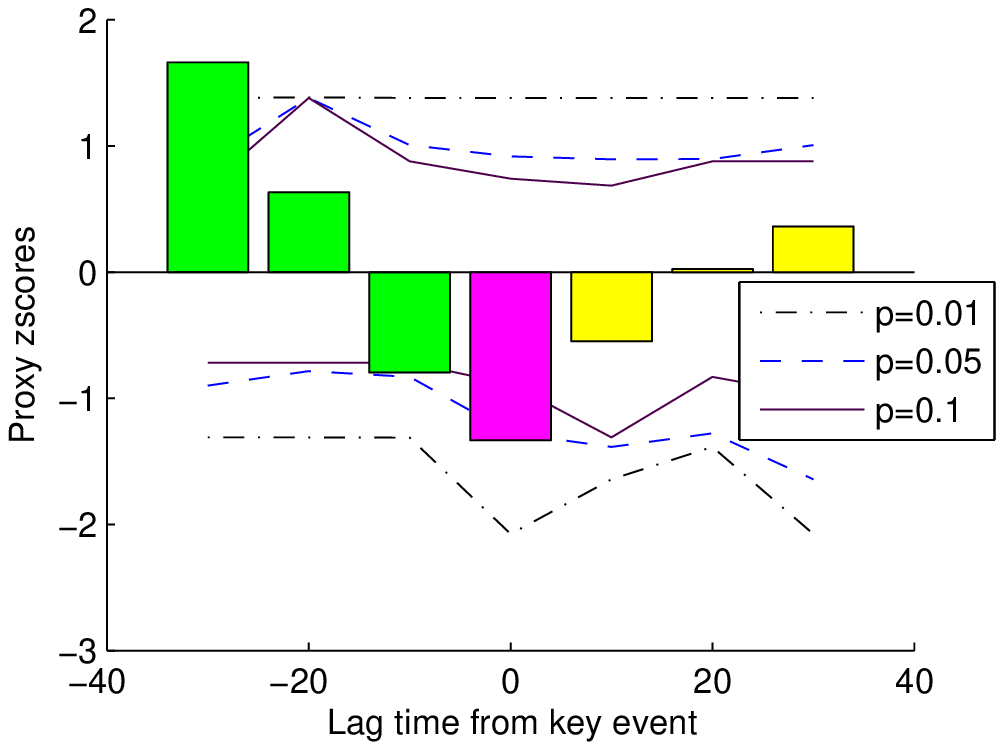}}
 \caption{\label{fig:SEA} The superposed epoch analysis showing proxy response of RSSN time series to "the maxima events" including 1989 grand maximum and "minima events" including 1685 the Maunder minimum which are spanned by the Suess/de Vries cycle. RSSN time series have very significant(p<0.01) rise at "the maxima events"(a) and significant(p<0.05) fall at "the minima events"(b). The green and yellow bars stand for pre- and post-"event" trends respectively and the central magenta bar for the zero epoch, i.e. at "the Suess/de Vries cycle maxima" or "minima events."}
\end{figure}

We have applied the superposed epoch analysis(SEA) on RSSN time series related to the "events" spanned by 207-year periodicity which are "maxima" including 1989 grand maximum and "minima" including 1685 the Maunder minimum. "Maxima" and "minima events" have the phase difference of 110 year, i.e. about half a cycle. We used Blarquez's SEA routine.\footnote{\url{http://paleoecologie.umontreal.ca/public/code/SEA.m}}\citep{Blarquez2010} Results are shown in Figure~\ref{fig:SEA}. Proxy response to the "maxima events" appears very significant at more than 99\% confidence level(p<0.01) during the events occurrence.(at a lag close to zero) Otherwise the response to the "minima events" is significant at 95\% confidence level which reasons why the periodicity of about 200-year can be more easily found in the lists of grand maxima rather than the lists of grand minima in \citet{Inceoglu2015, Inceoglu2016} and \citet{Usoskin2016}. Anyhow, the analysis shows that we can find the significant Suess/de Vries cycle in RSSN time series.

We have also a rough chi-squared test for the relation between the Suess/de Vries cycle and RSSN time series. Let $Y_d(i)$ be the $i$-th year of "maximum" expected with the Suess/de Vries cycle starting from 1989 and $Y_r(i)$ the year of the nearest maximum(though it may not be a grand maximum) in RSSN series. If we suppose the standard deviation of difference between them $\sigma_Y$ constantly, then the chi squared can be evaluated as 
\begin{align}
\chi^2=\sum_i \left(\frac{Y_r(i)-Y_d(i)}{\sigma_Y} \right)^2.
\end{align}
\citep[see an instance of the chi-squared test in][]{Jik-su2017}

From the chi-squared test $\sigma_Y$ is found to be more than 22.2 in $1\sigma$ confidence level(68\%), 19.6 in $2\sigma$(95\%) and 17.8 in $3\sigma$(99\%). It shows that we could find a RSSN grand maximum within 20 years from the expected year with the Suess/de Vries cycle. The deviation may be due to the intrinsic variance of period for the Suess/de Vries cycle or a maximum shift from superposition of other cycles.

Note that there are some double or multiple grand maxima which mean the absence of grand minimum between adjacent two grand maxima. The presence of multiple grand maxima would contradict to the fact that the grand minima have the periodicity of the Suess/de Vries cycle. If we are now living in a multiple grand maximum, then coming new grand minimum might be delayed rather than at the beginning of this century as some people expected.

\section{Discussion}\label{sec:disc}

We cannot exclude the possibilities that other cycles are deterministic which would be reasoned from the fact that power of peaks corresponding to longer cycles like the Hallstatt cycle are much more depressed in power spectrum of difference series than that of the original RSSN series. Suppose that original time series is a digitalization of a simple sinusoidal function:
\begin{align}
X(i)\sim X=A\cos\left(\frac{2\pi t}{T} \right).
\end{align}
Then its difference series should be corresponding to the derivation of the sinusoid function:
\begin{align}
\Delta X(i)\sim dX=\frac{2\pi}{T} A\cos\left(\frac{2\pi t}{T} \right)dt,
\end{align}
which showes that the longer a cycle is, the weaker amplitude it will have in Fourier spectrum of difference series. We emphasize that the autocorrelation  $R_{\Delta}$(Eq.~\eqref{eq:ac}) of difference series has the similar time variation and squarely proportional amplitudes as difference series $\Delta X(i)$ according to the Wiener-Khinchin's theorem. So the longer the cycle may be, the weaker power it will have in power spectrum of difference series.

We can explain the degradation rule of peaks in Figure~\ref{fig:samplogram} assuming that the cycles are deterministic sinusoidal. The calculation is long and tedious so we omit it here. However, we can say that power of a deterministic peak degrade approximately in inverse proportion to the square root of sampling interval, more exactly, sampling step, that is, the sampling interval divided by the time step of pre-down-sampled RSSN series.\citep[e.g. 10-year for RSSN of ][]{Wu2018}. This rule holds well especially in low frequencies, that is, much long cycles like the Hallstatt cycle and short sampling interval. 

The diagram for difference series(Figure~\ref{fig:samplogram}b) shows another fact about the behaviors of longer "cycles" at their Nyquist limit. As said above, the longer "cycles" are more depressed in power spectrum by differencing. While increasing sampling interval, their power would degrade more and more. In spite of that, you can find that "peaks" are soaring just before the Nyquist limit and decay rapidly beyond the limit(Figure~\ref{fig:samplogram}b). This implies that we have to examine the behavior of a cycle at its Nyquist limit. The deterministic cycle should have some special property at their Nyquist limit which will add some criteria to test whether it is a deterministic or not. Even though differencing may be a kind of "low-frequency filter," it cannot depress revival of cycles at their Nyquist limit. Such revival or soaring at Nyquist limit is much clearer in diagram for difference series(Figure~\ref{fig:samplogram}b) with greater contrast. We can explain this fact: If we consider a sampling interval which is Nyquist limit of a "cycle", this sampling interval is greater than the Nyquist limit of shorter cycles so that the shorter cycles already decayed much. The Nyquist sampling acts as a "high-frequency filter." On the other hand, the longer cycles are still depressed in that sampling by differencing. So, between both the "high-" and "low-frequency filters," the "cycle" will be dominated and its power will be upgraded even to 1.(Recall that if we take the autocorrelation as normalized covariance Eq.~\eqref{eq:ac}, then the power implies a relative fraction of energy of the cycle) We can see that after one "cycle" passed its Nyquist limit, the next longer "cycle" is revived.

Above discussions are related mainly to longer cycles, which, however, will add justification to argument that the Suess/de Vries cycle is deterministic because, as said above, if a random peak there is, it should have randomly variable or rapidly decaying power in increasing sampling interval rather than the deterministic degression as said above. But the power of the Suess/de Vries cycle of raw RSSN series decays in inverse proportion to the square root of sampling interval. And power of the cycle for difference series has a peak at its Nyquist sampling.

Another interesting phenomenon that we should note is the tracks of abnormal peaks in aliasing region between the Nyquist limit and Fourier limit.(Figure~\ref{fig:samplogram}) More interesting is that a track crosses the Nyquist limit of the 207-year cycle. We found that this track is due to the aliasing effect. The track in aliasing region which crosses the Nyquist limit of 207-year cycle is just of aliasing peak of this cycle. As we showed in Figure~\ref{fig:acps}(b), the aliasing peak is no more than a reflection of the true peak. 

We would like to call above diagrams(Figure~\ref{fig:samplogram}) the samplogram. The samplograms show how power of peaks changes in average down-sampling. And it will give some criteria(e.g. the behavior at Nyquist limit) to test for deterministic cycle. Remember that samplogram has one more dimension--sampling interval--than power spectrum. We can hardly say about random part of the signal. But if we have more criteria about the deterministic cycle, then we can differentiate the deterministic cycles from a stochastic process. We think that the samplogram will be an effective tool to extract the deterministic cycles from a stochastic process. 

\section{Conclusion}

In this paper we studied cycles appeared in old Korean records of naked-eye sunspot observation. Through the investigation we found about 200 year cycle. In comparison we investigated modern reconstructions of solar activity from cosmogenic radionuclides. In analyzing the reconstructed solar activity, a difficulty is that current methods are hard to differentiate deterministic cycles from the stochastic process of solar activity. We introduced the samplogram and showed that the Suess/de Vries cycle is stable in increasing sampling interval up to the Nyquist limit, that is, a half of the cycle. We argued that if it had a random peak then its peak would vary randomly and could not show such significant stability. Thus we conclude that the Suess/de Vries cycle is a deterministic cycle involved in stochastic and "non-stationary" solar activity. 

There is a possibility that the Suess/de Vries cycle would be introduced into composing reconstruction of solar activity indirectly, for example, through the geomagnetism correction. It means that the cycle might not be of solar activity, but of geomagnetism. However, we can exclude that possibility through historical records of naked-eye sunspot observations. Clearly, those data show 200-year cycle. So we argue that Suess/de Vries cycle is an intrinsic cycle of solar activity. 

We newly introduced the concept of the samplogram to analyze the solar activity. Further investigation of it is needed but we think it will be an effective tool in analysis of stochastic process.
Finally, we carried out the superposed epoch analysis and showed that the Suess/de Vries cycle has the significant correlation with time series of the reconstructed solar activity. And we showed the presence of multiple grand maxima which could give a suspect to an occurrence of the new grand minimum in the near future or, at least, forecast its delay.

\section*{Acknowledgements}
\addcontentsline{toc}{section}{Acknowledgements}

K. Chol-jun has been supported by \text{Kim Il Sung} University during investigation. K. Chol-jun thanks Jo Jong-won for giving Korean historical data and Mun Ui-ri for a help to English. And we are thankful anonymous referee for detailed and helpful comments.






 
\bsp	
\label{lastpage}
\end{document}